# Mbps Experimental Acoustic Through-Tissue Communications: MEAT-COMMS

Andrew Singer
Coordinated Science Laboratory
University of Illinois at Urbana Champaign
Urbana, IL USA
acsinger@illinois.edu

Michael Oelze and Anthony Podkowa
Beckman Institute
University of Illinois at Urbana Champaign
Urbana, IL USA
{oelze, tpodkow2}@illinois.edu

*Abstract*—Methods for digital, phase-coherent acoustic communication date to at least the work of Stojanjovic, et al [20], and the added robustness afforded by improved phase tracking and compensation of Johnson, et al [21]. This work explores the use of such methods for communications through tissue for potential biomedical applications, using the tremendous bandwidth available in commercial medical ultrasound transducers. While long-range ocean acoustic experiments have been at rates of under 100kbps, typically on the order of 1-10kbps, data rates in excess of 120Mb/s have been achieved over cm-scale distances in ultrasonic testbeds [19]. This paper describes experimental transmission of digital communication signals through samples of real pork tissue and beef liver, achieving data rates of 20-30Mbps, demonstrating the possibility of real-time video-rate data transmission through tissue for in-body ultrasonic communications with implanted medical devices.

*Keywords—ultrasonic, acoustic communications, tissue, through body.*

## I. INTRODUCTION

Remote monitoring of patients using wireless capabilities can be categorized as "on-body" monitoring and "in-body" monitoring [1-2]. On-body monitoring refers to sensors placed on the surface of the body while in-body monitoring refers to sensors placed within the body, i.e., implanted medical devices (IMDs). In the case of IMDs, transmission is characterized by low peak power and low duty cycle to reduce the potential for adverse bio-effects and to extend battery life. Other approaches seek to recharge small batteries in IMDs wirelessly by converting energy from a transmitted signal from an external device (as either electromagnetic or ultrasonic wave energy) into electro-chemical storage. Therefore, the use of wave propagation for communication and interaction with IMDs is an integral part of current and future device development.

To date, cardiac patients represent the largest segment of patients making use of wireless telemetry from IMDs. However, IMD wireless telemetry in the human body is expanding rapidly [1]. Applications include implanted pacemakers and defibrillators [13,14,15,16], glucose monitors and insulin pumps [17], intracranial pressure sensors [3], epilepsy control [4], ingestible cameras for imaging the digestive track, and many more applications. Therefore, the increased demand for these devices and the opening up of new applications for IMDs will continue to amplify the role of these devices for patient care and management of disease.

Currently, most IMDs use radio-frequency (RF) electromagnetic waves to communicate through the body. The Federal Communications Commission (FCC) regulates the bandwidths that can be used for RF electromagnetic wave propagation available to IMDs. For example, the Medical Implant Communication Services (MICS), renamed the Medical Device Radiocommunication Service (MDRS), designates frequencies of operation ranging from 401-406 MHz. The corresponding maximum bandwidth allowed is 300 kHz, which inherently limits the communication rates of these devices, and is reported in current devices to be limited to a maximum of 50 kb/s [1].

Beyond bandwidth restrictions, the main limitation for using RF electromagnetic waves in the body is loss of signal that occurs because of attenuation in the body [5]. Losses in soft tissues are comparable to losses in salt water, which is a major constituent of soft tissues and is a high loss medium for propagation of RF electromagnetic waves. Soft tissues each have their respective high loss dielectric properties which result in scattering and multipath of signals as well as loss. In order to overcome these losses, higher power must be used and this can result in heating of tissues due to absorption. For these reasons the output power of RF devices is limited to 25 µW. Furthermore, adverse bio-effects associated with radiation of electromagnetic waves in the body have not been studied in detail and long term biological effects of heating and non-thermal effects, such as purported increased risk of cancer, warrant additional study [6]. These perceived risks can be as important as the actual risks in deterring progress. These issues have impeded progress in developing intra-body wireless networks, allowing devices to communicate with each other through the body and with external devices.

In this paper, an alternate communication channel is explored for IMD communications with external devices, i.e., the acoustic (ultrasonic) communication channel. For underwater applications, RF electromagnetic communications has long since been supplanted by acoustic communication. Acoustic or ultrasonic communication is the preferred communication channel underwater because sound (pressure) waves exhibit dramatically lower losses than RF and can propagate tremendous distances for signals of modest bandwidth. For example, SONARs and acoustic modems with center frequencies of around 10 kHz can achieve distances of greater than 10 km. Similar to the case for underwater, an acoustic communication channel in the body also has the benefits of low loss compared to RF electromagnetic communications. For several decades, ultrasound has been used to provide images of the body and has amassed a stellar safety record among the imaging modalities. Perhaps equally important, acceptance of ultrasound as a safe and effective imaging modality is clear from its widespread use for imaging in utero. Compared to RF electromagnetic wave propagation, ultrasound absorption in tissues at clinical frequency ranges is orders of magnitude lower, resulting in a dramatically lower potential for tissue heating [7]. Clinical ultrasound transducers (center frequencies from 1-20 MHz) are often high bandwidth, i.e., up to 100%, which could translate to high data rates for communication in the body. Ultrasonic waves propagating in the body for communications would not face interference from external networks. Therefore, ultrasound offers a safe, high speed and low loss communication channel compared to conventional RF electromagnetic communications in the body.

The use of ultrasonic communications to control and monitor IMDs is not new. Several researchers have proposed ultrasonic communication with IMDs and have conducted some preliminary work to show feasibility [8,9,10,11]. Santagati and Melodia developed a prototype intra-body sensor network using ultrasonic transducers and demonstrated in a tissue-mimicking phantom the ability to communicate ultrasonically with a data rate of 347 kbps [12]. In that work, an FPGA was programmed to implement an ultrasonic wideband technique with some resilience to multipath and a medium access control layer protocol [12]. In a different study, the ultrasonic communications channel was used not only to send information through a tissue-type channel (water), but the ability to power devices remotely through the ultrasound communication channel was demonstrated using ultrasonic waves [11]. Therefore, the ultrasonic communication channel has demonstrated the potential to be used for communicating and powering of IMDs in the body.

While some physical considerations of ultrasonic communications through tissues have been considered and some practical guidelines established, the ultrasonic communication channel in the body has not been fully characterized and as a result the full potential for high speed communications using ultrasound has not been realized. We demonstrate that improved signal processing techniques can provide even higher data rates with low error rates (>10 Mbps) through tissues at frequencies that would allow propagation through the body (< 10 MHz). These data rates are sufficient to allow real-time streaming of high definition video and to operate and control small devices within the body. For example, standard definition streaming of video requires 1.75 Mbps while high definition video streaming starts at 3.6 Mbps. Therefore, by communicating at rates up to 10 Mbps using ultrasound, we envision the ability to not only control IMDs in the body but to provide live streaming of HD video from devices inside the body. One can imagine a device that is swallowed for the purposes of imaging the digestive tract but with the capability for the HD video to be continuously streamed live to an external screen and the orientation of the device controlled wirelessly and externally by the physician.

In this paper, we provide results from ultrasonic communications experiments through tissue and validate the ability to achieve high data rates capable of real-time HD video streaming and remote control of tissue embedded devices. We have demonstrated the ability to transmit data at 120 Mbps [19] through water using a 20-MHz transducer and 20-MHz bandwidth. Specifically, in our experiments we used a 20-MHz f/3 single-element transducer with a -10-dB bandwidth of approximately 20 MHz to send information-bearing signals. The focus of the transducer was 1.9 cm and a needle hydrophone (HPM075, Precision Acoustics, Dorchester, UK) was used to record the transmitted signals. The hydrophone was broadband and covered the bandwidth of the transmitter. The transducer and hydrophone were placed in a tank filled with degassed water and faced each other at a distance of 1.9 cm. A picture of the experimental setup is shown in Fig. 1. A 64-QAM signal was generated in Matlab and uploaded to an arbitrary waveform generator (W1281A, Tabor Electronics, Tel Hanan, Israel). The QAM signal was preceded by superimposed up/down hyperbolic chirps for synchronization and to initialize the receiver for Doppler effects due to platform motion. Using this setup, a data rate of 120 Mbps was achieved over a distance of 1.9 cm. The raw equalizer output BER was about 2E-2 and can be made error-free with about 15% forward error correction (FEC) overhead. To our knowledge this result is at least 100x higher than any reported underwater acoustic communication experiment to date.

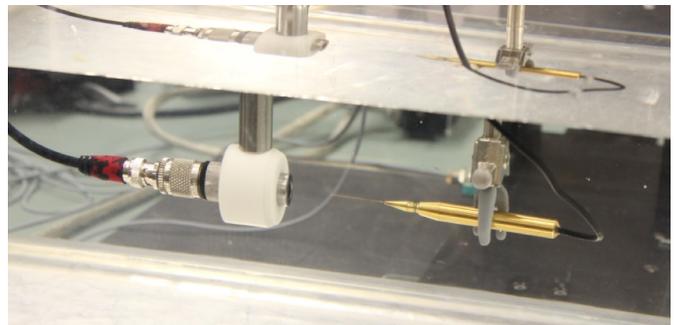

Figure 1. Experimental setup using a 20-MHz single-element transducer to send a 64-symbol QAM signal to a hydrophone.

## II. SIGNAL PROCESSING MODEL

### A. Acoustic Communications (ACOMMS) Methods

The most spectrally efficient ACOMMS methods reported have employed passband quadrature amplitude modulation (QAM) for coded communications [19]. Equalization and tracking methods [18-21] have shown that receivers for such QAM signals can be resilient to Doppler and multipath reverberation and scattering [18]. In the experiments described in this paper, passband QAM signals were constructed from baseband k-ary QAM signals, $x[k] \in S$, where $S$ is a $2^k$-ary symbol alphabet, for $k = \{2,3,4,6\}$, resulting in QPSK, 8PSK, 16QAM and 64QAM symbols, respectively.

The pass band signal comprised a 13-symbol Barker sequence {1,-1,1,-1,1,1,-1,-1,1,1,1,1,1}, or a 10 microsecond quadratic chirp from $\frac{-f_b}{2}$ to $\frac{f_b}{2}$, where $f_b = 1/T_s$ is the symbol rate, at center frequency $f_c$, followed by a 1 msec guard interval, N=50,000 QAM symbols and 1msec guard before a subsequent transmission. The pass band signal can be written as $x(t) = \mathcal{Re}\{\sum_{k=0}^{N} x[k]p(t - kT_s)e^{j2\pi f_c t}\}$, where $p(t)$ is a raised cosine filter with roll-off factor 0.8. An example waveform and its spectrogram are shown in Figure 2 below, for $f_b = 5MHz$, $f_c = 4MHz$, $k = 6$.

In our experiments we used a matching pair of 5-MHz f/3 single-element transducers (Valpey Fisher, IL0506HR) operating in a pitch-catch configuration. The transducers had a

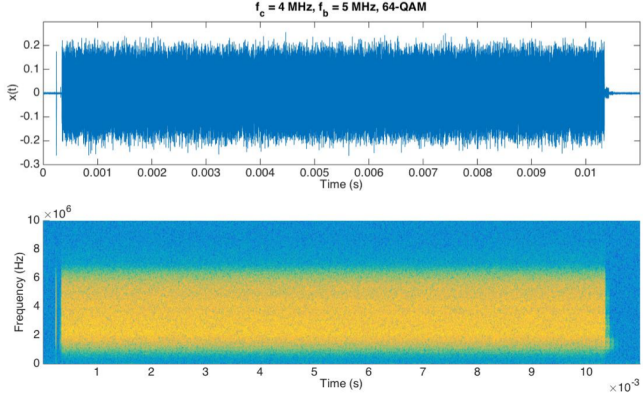

Figure 2. Example transmit waveform and its spectrogram with a 5MHz symbol rate, 4MHz center frequency, and 64QAM modulation alphabet. A Barker synchronization sequence is used for this packet.

-10-dB bandwidth of approximately 5 MHz to send/receive information-bearing signals. The transducers had a nominal focus of 5.72 cm and a 1.92 cm diameter. The transducers were placed in a tank filled with degassed water and faced each other at a distance of 5.86 cm. A picture of the experimental setup is shown in Fig. 3. The QAM signals were generated in Matlab and uploaded to an arbitrary waveform generator (Tabor, W1281A) and used to drive the transducers via an 55-dB amplifier (ENI A150).

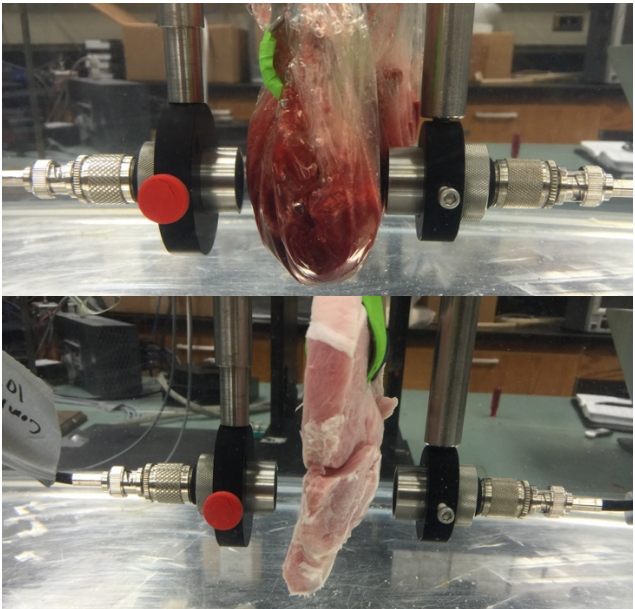

Figure 3. Experimental setup using two 5-MHz transducers to send pass band QAM communication signals through beef liver (top) and pork loin (bottom).

For both the pork loin and beef liver samples, the samples were suspended into the transmit/receive acoustic signal path, and signals were transmitted and captured by sending 10 snapshots (packets) of 50,000 samples (10,000 training / 40,000 decision-directed) using a fractionally-spaced (2 samples per symbol) decision feedback equalizer with up to 40 taps in the feed forward section, and 40 taps in the feedback section. The equalizer was operated in decision-directed mode using the recursive-least squares algorithm with exponential forgetting factor of 0.995, along with phase-tracking as developed in [21], using a second-order phase-locked loop with numerator polynomial [0.0011 -0.001 0] and denominator polynomial [1 -2 1]. The pork loin was suspended directly in the signal path, while the beef liver was suspended within a saran-wrapped enclosure, as shown in Fig. 3. For each of the signal parameters shown in Table I, signals were generated, transmitted, recorded, and decoded. Shown in the table, the bit error rates are given as < 1E-4, since the packets were of length 1E+4 and transmitted error free after training. The data point from the last row in the table was not of sufficient signal-to-noise ratio to permit decoding at the time of this paper.

TABLE I. EXPERIMENTAL TRANSMISSIONS

| Channel Type | Modulation Parameters and Results | | | | |
|---|---|---|---|---|---|
| | Format | $f_c$ | $f_b$ | Data Rate | BER |
| Pork Loin | QPSK | 5MHz | 2.5MHz | 5Mb/s | <1E-4 |
| Pork Loin | 16QAM | 5MHz | 2.5MHz | 10Mb/s | <1E-4 |
| Pork Loin | 64QAM | 5MHz | 2.5MHz | 15Mb/s | <1E-4 |
| Pork Loin | 16QAM | 5MHz | 5MHz | 20Mb/s | <1E-4 |
| Pork Loin | 64QAM | 5MHz | 5MHz | 30Mb/s | <1E-4 |
| Pork Loin | 64QAM | 4MHz | 5MHz | 30Mb/s | <1E-4 |
| Beef Liver | QPSK | 5MHz | 2.5MHz | 5Mb/s | <1E-4 |
| Beef Liver | 64QAM | 5MHz | 2.5MHz | 15Mb/s | <1E-4 |
| Beef Liver | QPSK | 5MHz | 5MHz | 10Mb/s | <1E-4 |
| Beef Liver | 16QAM | 5MHz | 5MHz | 20Mb/s | <1E-4 |
| Beef Liver | 64QAM | 5MHz | 5MHz | 30Mb/s | * |

Table 2. Experimental data collected in ultrasonic experiments. QAM Sets comprise 4QAM(QPSK), 16QAM, or 64QAM, center frequency Fc, Symbol Rate Fb, Synch Pulse is either Barker or 10us quadratic Chirp, Data Rate represents the raw channel data rate before FEC, and Error Rate is an estimate of uncoded BER at the output of the equalizer.

For the third row in Table I, a 64QAM, 5-MHz center frequency signal with a 2.5-MHz symbol rate was transmitted, and the resulting mean-squared error (MSE) is shown in Fig. 4 (top) along with the receive signal constellation after equalization. The fourth row of Table I, included a 5-MHz symbol rate, and 16QAM signaling, resulted in the MSE and signal constellation shown in the second from the top in Fig. 4. The sixth row of Table I, corresponding to a 4-MHz center frequency, 5-MHz symbol rate, with 64QAM is shown in the third from the top in Fig. 4. The eighth row in Table I, corresponding to a 2.5-MHz symbol rate, 5-MHz center frequency, and 64QAM signaling is depicted in the bottom of Fig. 4.

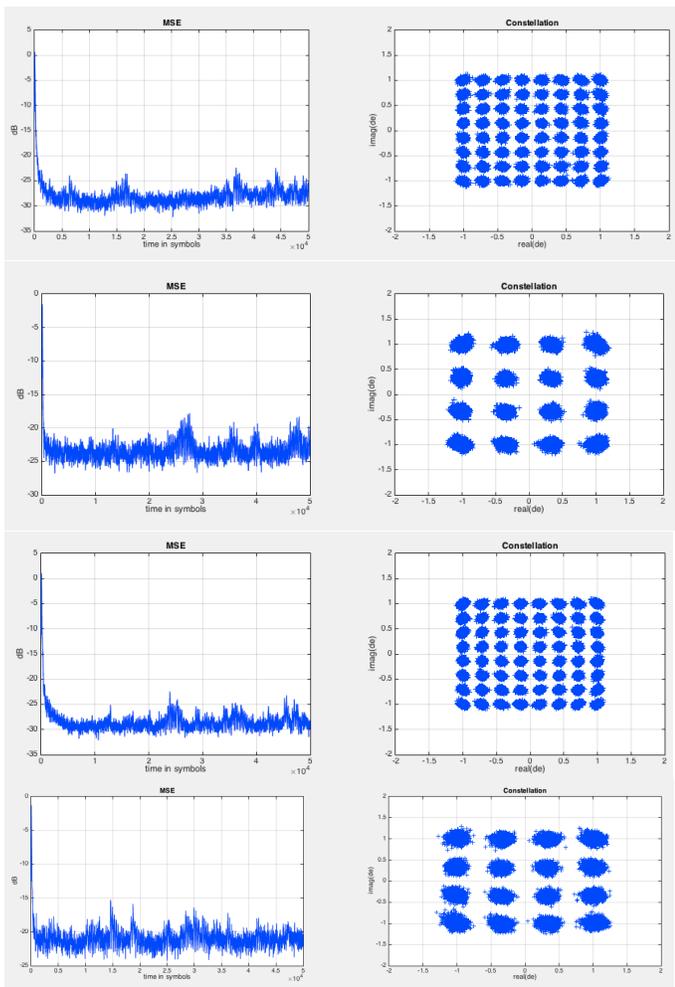

Figure 4. (top) Table I row 3: 64 QAM, 5MHz center frequency signal with a 2.5MHz symbol rate was transmitted through pork loin (PL), and the resulting mean-squared error (MSE) and receive signal constellation after equalization. (second) Table I row 4: 5MHz center frequency, 5MHz symbol rate, 16QAM through PL, (third) Table I row 8: 5MHz center frequency, 2.5MHz symbol rate, 64QAM through beef liver (BL), and (bottom) Table I row 10: 5MHz center frequency, 5MHz symbol rate rate, 16QAM through BL.

ACKNOWLEDGMENT

This work was supported in part by Systems on Nanoscale Information fabriCs (SONIC), one of the six SRC STARnet Centers, sponsored by MARCO and DARPA. This work was supported in part by the department of the Navy, Office of Naval Research, under grants ONR MURI N00014-07-1-0738 and ONR N00014-07-1-0311.